# Current redistribution model of anomalous resistance behaviour in superconductor-topological insulator heterostructures

Abhirami S.[1,3], Edward Prabu Amaladass[1,3], Prashant Sharma[2,3], Vinod K.[1,3], Thanikaiarasu A.V.[1], Awadhesh Mani[1,3*]

[1]*Materials Science Group, Indira Gandhi Center for Atomic Research, Kalpakkam-603102, Tamil Nadu, India*

[2]*Reactor Design and Technology Group, Indira Gandhi Center for Atomic Research, Kalpakkam-603102, Tamil Nadu, India*

[3]*Homi Bhabha National Institute, Training school complex, Anushaktinagar, Mumbai-400094, India*

*Corresponding author email: mani@igcar.gov.in

## Abstract

Anomalous resistance upturn and downturn have been observed on the topological insulator (TI) surface in superconductor-TI (NbN-$Bi_{1.95}Sb_{0.05}Se_3$) heterostructures at ~ mm length scales away from the interface. Magnetotransport measurements were performed to verify that the anomaly is caused due to the superconducting transition of the NbN layer. The possibility of long range superconducting proximity effect due to the spin-polarized TI surface state was ruled out due to the observation of similar anomaly in NbN-Au and NbN-Al heterostructures. It was discovered that the unusual resistance jumps were caused due to current redistribution at the superconductor-TI interface on account of the geometry effects. Results obtained from finite element analysis using COMSOL package has validated the proposed current redistribution (CRD) model of long range resistance anomalies in superconductor-TI and superconductor-metal heterostructures.

## 1. Introduction

Fu and Kane proposed that *p*-wave superconductivity, or topological superconductivity, is induced on the surface of a topological insulator (TI) due to proximity with an *s*-wave superconductor [1]. Such a *p*-wave superconducting state has been predicted to host Majorana zero modes inside the Abrikosov vortex cores [1,2]. Majorana modes are potential candidates as qubits for fault tolerant quantum computation owing to the fact that they obey non-Abelian statistics [3]. Over the years, many studies on superconductor-TI devices have successfully observed an induced superconducting state in the TI [4–12]. However, proving the unconventional nature (*p*-wave nature) of the induced pairing has been a challenge. Proximity-induced superconducting gap in the TI has been observed through techniques like dynamic conductance [13–15], scanning tunneling spectroscopy [16,17], and point contact spectroscopy [18]. Some studies have reported a double transition in temperature dependent resistivity (R-T) measurements [8,19,20]. Zero bias conductance peak (ZBCP) in the dynamic conductance spectrum or scanning tunneling spectroscopy spectrum has been reported in a few studies [13,14,21], and some of them have attributed the ZBCP to induced *p*-wave superconductivity [13]. A Few reports have inferred the presence of Majorana zero modes in the induced superconducting state [13,16,21–23]. Interestingly, induced superconductivity has been observed in TI up to greater lengths than the superconducting coherence length, particularly in Josephson junction devices [6,8,11,12,24]. Such a long range proximity has been speculated to be due to a combination of the spin-triplet pairing on the TI surface state and the selective coupling of the induced superconductivity to the ballistic surface states of the TI rather than to the bulk [4,6,11]. In some superconductor-TI devices, a small drop in the temperature-dependent resistance of the TI has been observed at the superconducting transition temperature ($T_c$) of the superconductor [19,20,25]. Even though the resistance does not approach 0 Ω and the length scale is far greater than expected, such a drop has often been reported as a signature of the induced long range superconductivity due to the spin-triplet nature of the TI surface state. Ironically, other studies [26–28] have reported an upturn in resistance at $T_c$ in similar superconductor-TI heterostructures at comparable length scales and ascribed this phenomenon also to the spin-triplet pairing of the TI surface state, however with a key difference in the explanation. It has been suggested that the spin non-degenerate TI surface state being incompatible with the *s*-wave pairing of the superconductor, causes the breaking of Cooper pairs. This is followed by a spin flip event which causes an increase in the resistance at $T_c$. Quantum confinement has also been cited as the reason for resistance upturn at $T_c$ in some devices [29]. Clearly, the agreement to the proposed origins of such contradictory resistance upturn and downturn signatures in superconductor-TI devices is not unequivocal. It is possible that the ambiguity in the behaviour is due to difference in geometry or nanostructure of the heterostructures, or the nature of the interface in the different studies. However, in one work [20] on In-Bi$_2$Te$_3$-In heterostructures, where multiple samples of similar geometry, nanostructure, and interface were prepared and studied, some samples showed upturn in resistance at T$_c$ while other samples showed downturn.

Such ambiguities necessitate a detailed investigation to bring out a unifying origin for the observed long range resistance upturns and downturns in superconductor-TI heterostructures. With this objective, we have fabricated NbN-BSS (Bi$_{1.95}$Sb$_{0.05}$Se$_3$) heterostructures and studied the temperature and magnetic field dependent resistivity behaviour of the TI region near the superconductor-TI interface. Interestingly, both the upturn effect and downturn effect have been observed on opposite edges of the device at ~ mm length separations from the interface. We have observed similar anomalous resistance upturn and downturn in superconductor-metal heterostructures, thus proving that the anomaly does not originate in the spin-polarized TI surface state. This manuscript demonstrates that the observed anomaly is due to current redistribution effects at the interface of the superconductor-metal or superconductor-TI device. Results obtained from finite element analysis based simulation of the current distribution across the superconducting transition of the heterostructure has conclusively proved the validity of the proposed current redistribution (CRD) model.

## 2. Heterostructure fabrication and measurement configuration

Superconductor-TI heterostructures have been fabricated in the step geometry. The TI of choice, which forms the first layer of the heterostructure, is Sb-doped Bi$_2$Se$_3$ (BSS) thin film. The BSS films of ~100 nm thickness have been grown in a pulsed laser deposition system on Si/SiO$_2$ substrates of dimensions 10 mm × 4 mm at 350 °C substrate temperature and in 5×10$^{-1}$ mbar argon environment [30–32]. The TI film was masked using a thin Aluminium (Al) shadow mask such that an area of 5 mm × 4 mm is exposed for the deposition of the ~100 nm thick niobium nitride (NbN) superconducting layer. NbN has been deposited using DC magnetron sputtering at 350 W power, 319 V bias voltage, and 1.1 A current for a duration of 10 min. The argon and nitrogen flow were maintained at 33 standard cubic centimeters per minute (SCCM) and 7 SCCM, respectively. The schematic of the fabrication of the heterostructure is depicted in Figure 1.

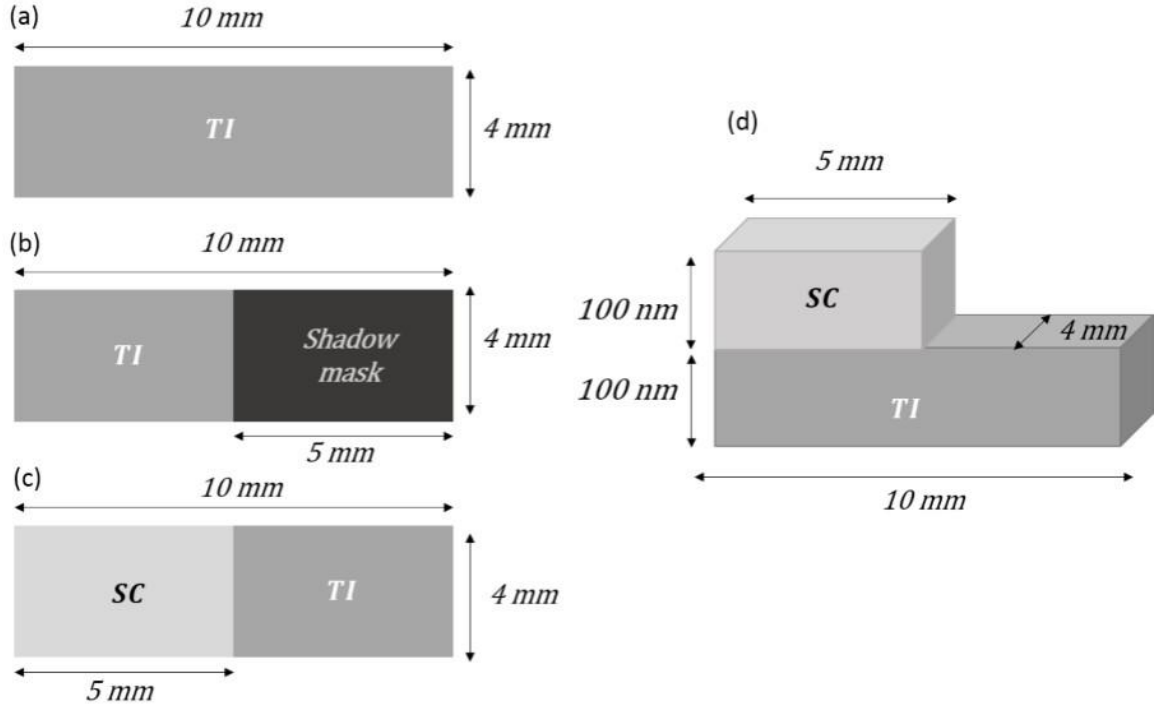

*Figure 1. Top view of the (a) Sb-Bi$_2$Se$_3$ (BSS) topological insulator (TI) thin film grown on Si/SiO$_2$ substrate using PLD technique, (b) TI thin film which is partially shadow masked for superconductor (SC) deposition, (c) superconductor-topological insulator heterostructure; the SC is deposited on the unmasked region of the TI. (d) Lateral view (3D view) of the superconductor-topological insulator heterostructure.*

The temperature-dependent resistance of the heterostructure has been measured in the linear geometry. A schematic of the electrical measurement configuration is depicted in Figure 2. Constant DC current of 1 mA is sourced across the heterostructure and the voltage drop on the SC and on the TI have been simultaneously measured. As shown in the schematic shown in Figure 2 the voltage drop on the TI has been measured at different regions on the TI. The width of the heterostructure is represented along the $X$-axis and the separation from the superconductor-TI junction edge is represented along the $Y$-axis. The $X = 0$ line of the Cartesian plane is arbitrarily made to lie halfway (2 mm) along the width of the sample. The SC-TI junction edge lies on the $Y = 0$ line and shall henceforth be referred to as the interface. The masked region of the TI during SC deposition, i.e., the bare-TI layer, lies on the positive $Y$-axis. The $(X, Y) = (0,0)$ origin point is marked as *(0,0)* in Figure 2 (b). The current is sourced along the $Y$ direction. The voltage drop values are measured at ~ mm length scales away from the junction. In Figure 2, $V_{X+}$ represents the voltage drop along the current direction in the (+X, +Y) region of the TI, while $V_{X-}$ represents the voltage drop in the (-X, +Y) region of the TI. The X co-ordinate and Y co-ordinate values are in units of mm, in accordance with the dimensions of the sample. All the electrical contacts were made manually with copper wires secured on the sample's surface using highly conducting silver paint.

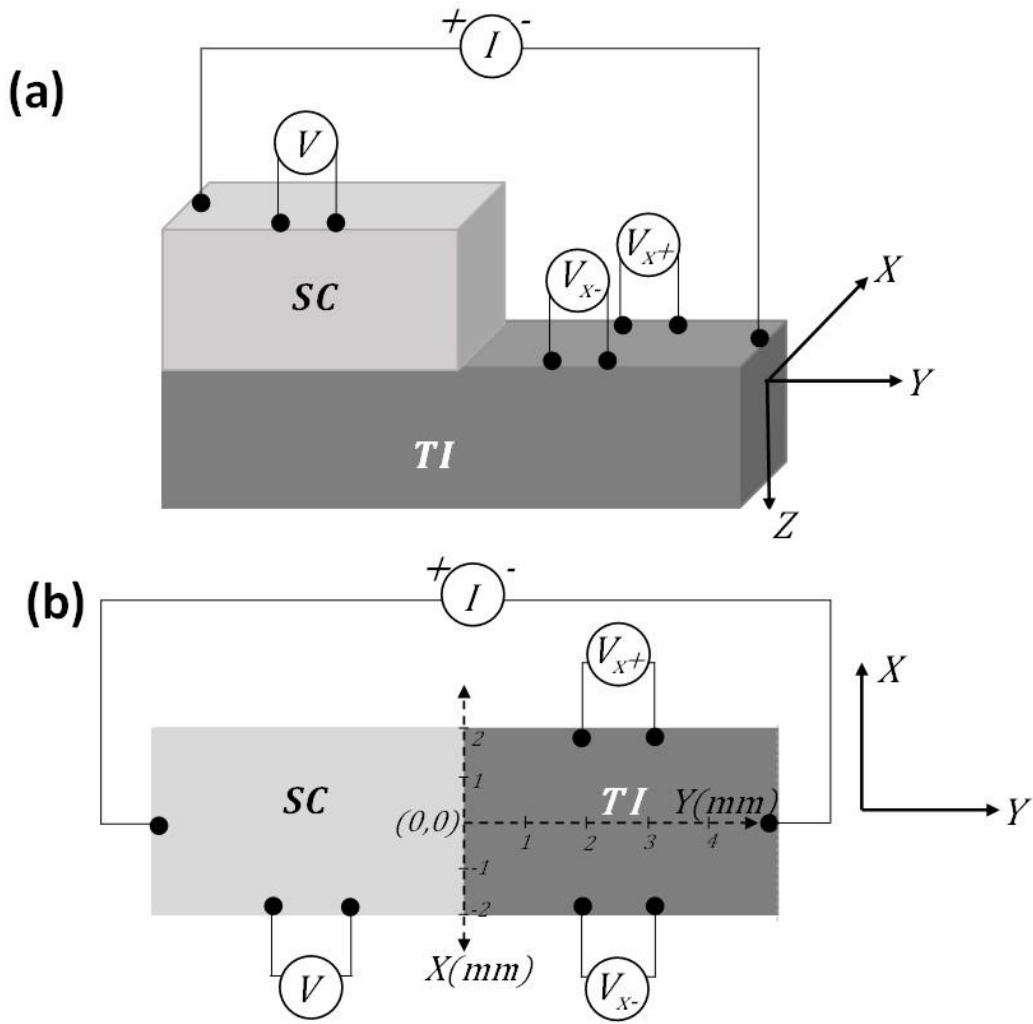

*Figure 2. (a) Lateral view and (b) Top view of the schematic representation of the electrical measurement configuration on the superconductor-topological insulator heterostructure.*

## 3. Results

Temperature dependent resistance measurements were carried out between 4 K and 15 K on the NbN-BSS bilayer and on different regions on the bare-TI layer of the heterostructure. It has been shown in previous studies on NbN-BSS heterostructures [32] that the superconducting transition temperature of ~100 nm NbN thin film deposited on ~100 nm BSS does not change significantly from that of the reference NbN thin film deposited on Si/SiO$_2$. Hence the T$_c$ of the bilayer will henceforth be referred to as the T$_c$ of the superconducting layer.

A sharp upturn in the temperature dependent $V_{X+}$ curve of the bare-TI was observed during the cooling cycle at the T$_c$ of NbN. Simultaneously, a downturn was observed in the $V_{X-}$ curve of the TI. The R-T curves of BSS at $V_{X+}$ and $V_{X-}$ are shown in Figure 3 (a), superimposed over the R-T curve of the NbN-BSS bilayer. This is an anomalous feature that needed further investigation. It was subsequently found that magnetic field had a similar effect as that of temperature on the resistance of the TI at the superconducting transition. At fixed temperatures, the resistance at different regions on the heterostructures was measured as a function of a perpendicular magnetic field varying between ±15 T. As the magnitude of the field was decreased from 15 T, $V_{X+}$ was found to exhibit an upturn at the upper critical field H$_{c2}$ of NbN. Parallelly, the magnetoresistance curve of $V_{X-}$ of the TI is found to exhibit a downturn at H$_{c2}$. This behaviour has similarities to the R-T behaviour shown in Figure 3 (a) and hence the anomalies in R-T and MR studies must originate from the same physical phenomenon. The observation of the opposing modifications in in the resistance of the two edges at T$_c$ suggests that when NbN goes into the superconducting state, the bare-TI layer of the heterostructure is immediately divided into a high resistance region (+X,+Y) and a low resistance region (-X,+Y) along the width. The magnetoresistance curves of $V_{X+}$ and $V_{X-}$ at 10 K are shown in Figure 3 (b).

The anomalous R-T jumps of $V_{X+}$ and $V_{X-}$ were recorded at different fixed magnetic fields in order to study the effect of the shifting T$_c$ of NbN on the anomaly. It was observed that increasing the magnitude of the magnetic field caused the upturn/downturn to shift to lower temperatures such that the feature always appears at the T$_c$ of NbN, which decreases to lower temperatures in the presence of applied magnetic field. This phenomenon is depicted in the R-T curves in Figure 4. A lowering in the magnitude of the upturn/downturn is observed with increasing field, indicating that the feature is strongest at B = 0 T where the superconductivity is strongest. The upturn/downturn feature was also observed to broaden with increasing magnetic field in correspondence with the broadening of the superconducting transition of NbN. Hence, it is unambiguously inferred that the anomalous upturn/downturn in the bare BSS layer near NbN-BSS interface arises due to the superconducting transition of the NbN-BSS bilayer.

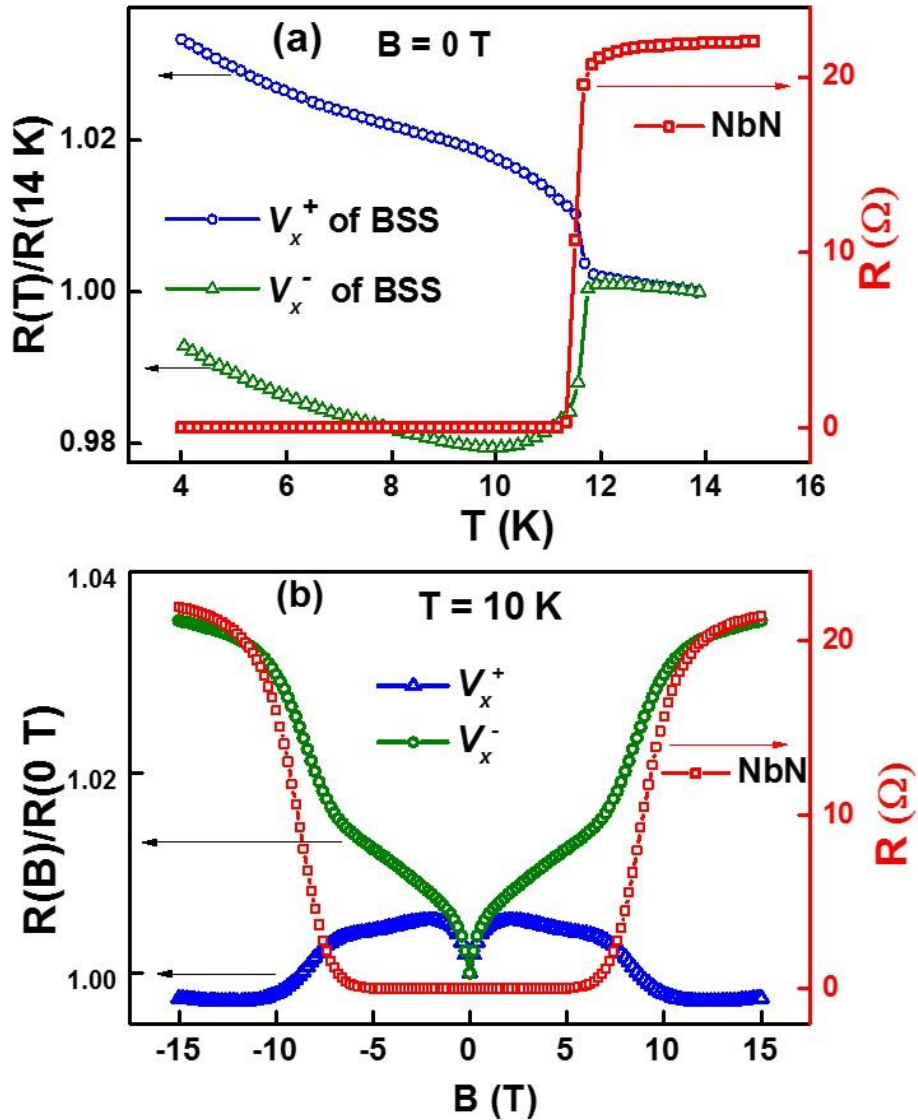

*Figure 3. (a) Temperature-dependent resistance (normalized by the resistance at 14 K) at 0 T and (b) Magnetic field-dependent resistance (normalized by the resistance at 0 T) at 10 K of the bare-TI layer of the NbN-BSS heterostructure measured on opposite edges of the width of the heterostructure ($V_{X+}$ and $V_{X-}$). In (a) the temperature-dependent resistance curve of NbN superconductor layer (deposited partially over the TI) is superimposed over the resistance curves of the TI.*

Such magnetic field dependent shifting in the resistance upturn/downturn signatures have been observed by Afzal *et al.* [27] in $MoTe_2$ pellets contacted with indium (In) electrodes and by Yadav *et al.* [25] in $Bi_2Se_3$ flakes contacted with tungsten (W) electrodes.

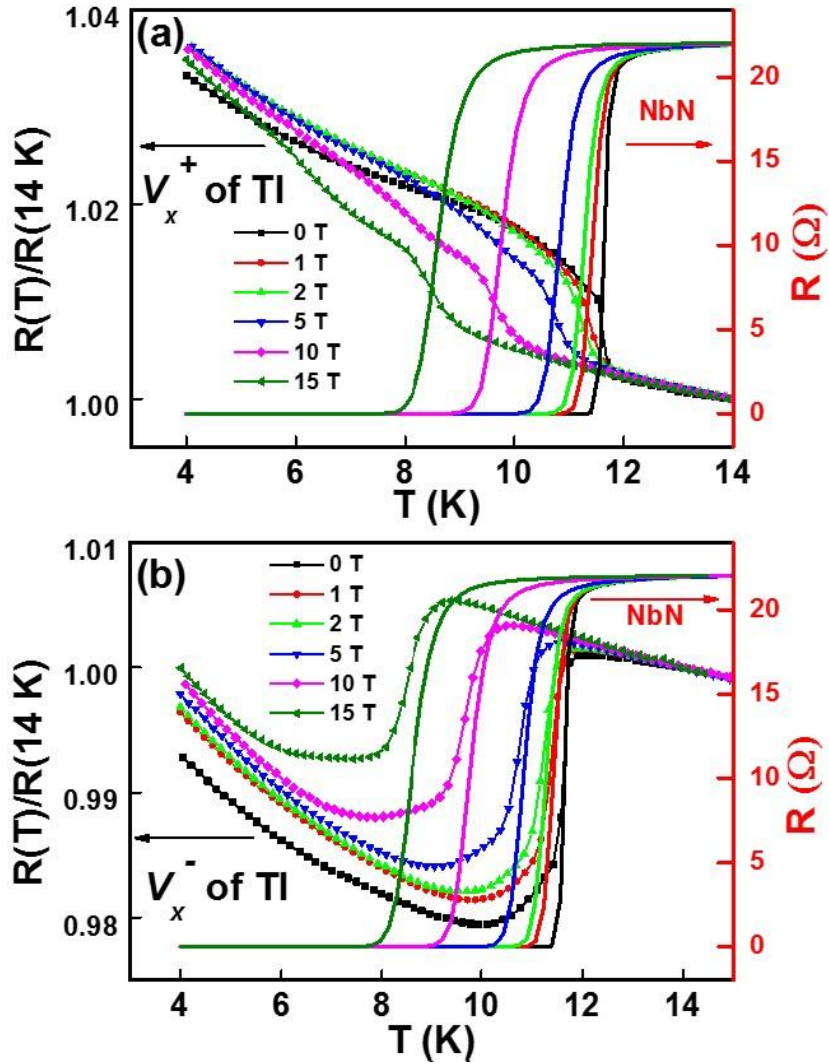

*Figure 4. Temperature-dependent resistance (normalized by the resistance at 14 K) of bare-TI layer of the NbN-BSS heterostructure measured on opposite edges of the width of the heterostructure (a) $V_X+$ and (b) $V_X-$, at different fixed magnetic fields between 0 T and 15 T. In both (a) and (b) the R-T curves of the bare-TI are superimposed with the R-T curves of the NbN-BSS bilayer depicting the lowering of $T_c$ of NbN with increasing magnetic field and a corresponding lowering in the upturn/downturn temperature. The symbols represent the R-T data points of the bare-TI and the solid lines represent that of NbN-BSS bilayer.*

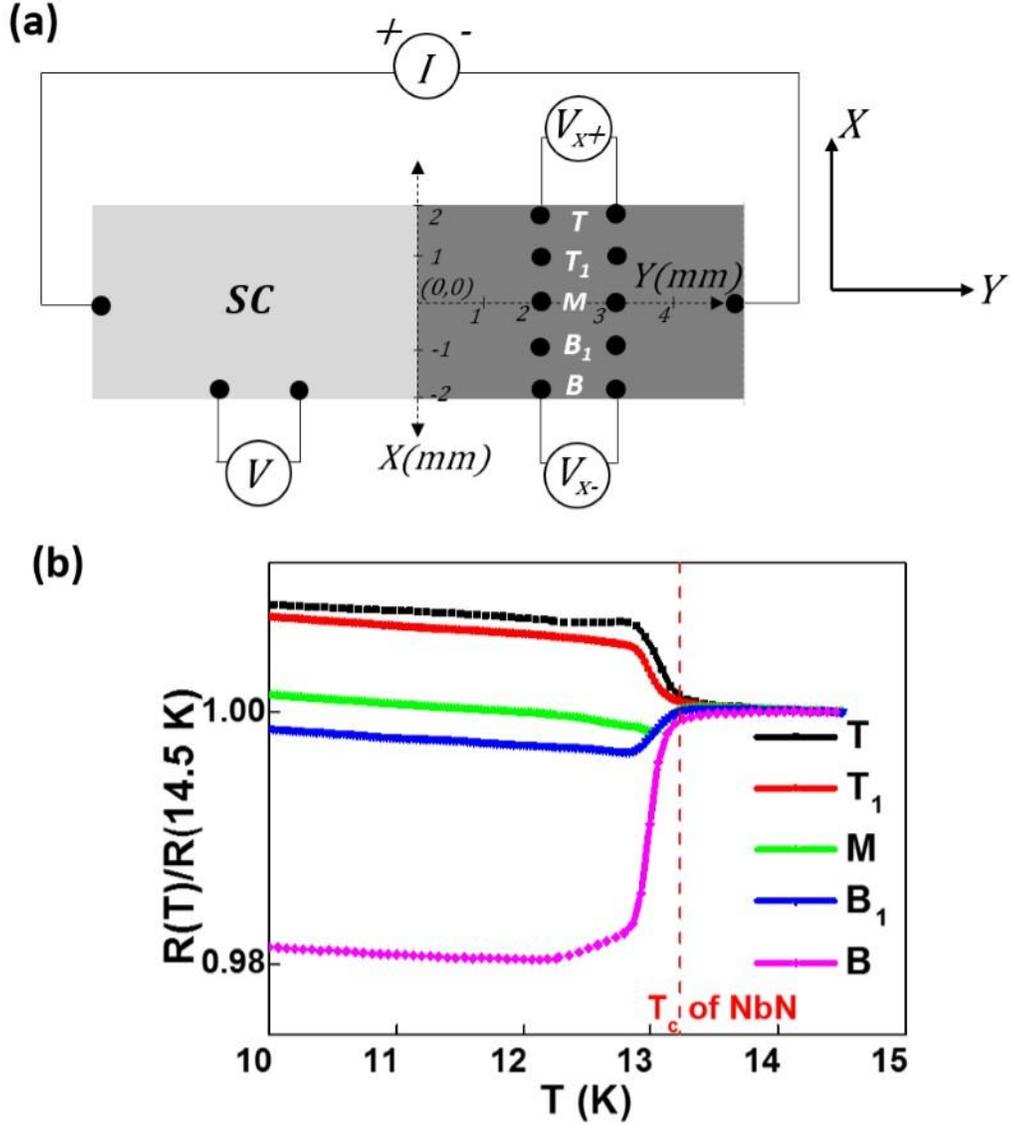

*Figure 5. (a) Schematic representation of measurement configuration for the study of the R-T anomaly at different regions across the width of the bare-TI later of the NbN-BSS heterostructure. T represents the voltage pair at X ~ +2, $T_1$ at X ~ +1, M at X ~ 0, $B_1$ at X ~ -1, and B at X ~ -2. (b) Temperature-dependent resistance (normalized by the resistance at 14.5 K) of bare-TI layer measured on five different pairs of voltage leads fabricated at different x values across the width of the sample.*

The contrasting behaviour of the anomaly on the opposite edges of the width of the heterostructure indicated that the anomalous jump should vanish at the $X = 0$ line, i.e., along the middle of the width of the TI. In order to confirm this notion, five pairs of voltage leads were made across the width on the bare-TI layer of the heterostructure. A schematic representation of the measurement configuration is given in Figure 5 (a). *T* represents the voltage pair at $X \sim +2$, $T_1$ at

$X \sim +1$, $M$ at $X \sim 0$, $B_1$ at $X \sim -1$, and $B$ at $X \sim -2$. The R-T curves obtained from the voltage drops at different values of $X$ are shown in Figure 5 (b). The voltage pair $T$ is seen to show an upturn of highest amplitude. With decreasing $X$ (towards $X = 0$) in the $(+X,+Y)$ quadrant, the amplitude of the upturn ($T_1$ pair) is seen to decrease. It is found to become insignificant along the $X = 0$ line ($M$ pair). With increasing magnitude of $X$ in the $(-X,+Y)$ quadrant ($B_1$ pair), the upturn is found to flip to a downturn and further grow in amplitude ($B$ pair). Hence the anomalous jump in resistance on the TI at the $T_c$ of the superconductor is found to be strongest at the edges of the width and seen to vanish at the middle. The electrical contacts are manually made and hence may not be equally spaced along width. It is possible to presume that the difference in amplitude of the $T$ pair and the $B$ pair are due to possible errors in spacing between the manually positioned leads.

Superconducting proximity effects are confined to the length scale of the superconducting coherence length, which is typically of the order of a few nm. The coherence length of bulk NbN is reported in literature to be ~ 7 nm [33]. The 100 nm NbN film deposited on the BSS layer in this work has been found to possess a coherence length of ~ 3.5 nm [32]. Since the length scales at which the voltage drops are measured in the present study (mm scales) are six orders of magnitude higher than that of the coherence length of NbN, it is improbable that the resistance downturns/upturns observed in the TI layer are induced by conventional superconductor-metal proximity effects. However long range proximity in superconductor-TI heterostructure has been reported in literature for ~ μm length scales due to the unique spin-polarized surface states of the TI [6,8,11,12,24].

In order to find if the anomalous jump is unique to TI materials, the TI layer in the superconductor-TI heterostructure was replaced with an ordinary metal (gold (Au)) and R-T measurements were carried out in the NbN-Au heterostructures within 4 K and 15 K at 0 T and at non-zero magnetic fields within ±15 T. It was observed that the anomalous jump in resistance observed in NbN-BSS heterostructure is also present in the NbN-Au heterostructure, as shown in Figure 6 (a).

A large spin-orbit coupling strength is an essential feature of a TI. The spin-momentum locking of the surface is created by the spin-orbit coupling and further protected by the time-reversal symmetry in a 3D-TI [34]. The Au metal studied to verify the effect of the superconductor on metals is a high Z element and possesses a large spin-orbit coupling as well, although smaller than that of $Bi_2Se_3$. Hence it is essential to verify the results with a low Z metal like Aluminum

(Al), which has poor spin-orbit coupling strength. The measurements were repeated in heterostructures of NbN-Al, where the Al replaces the BSS layer (or the Au layer). R-T measurements revealed the presence of resistance jumps at the $T_c$ of NbN in the bare-Al layer of the NbN-Al heterostructure, confirming that the resistance anomaly is a feature common to all superconductor-metal heterostructures.

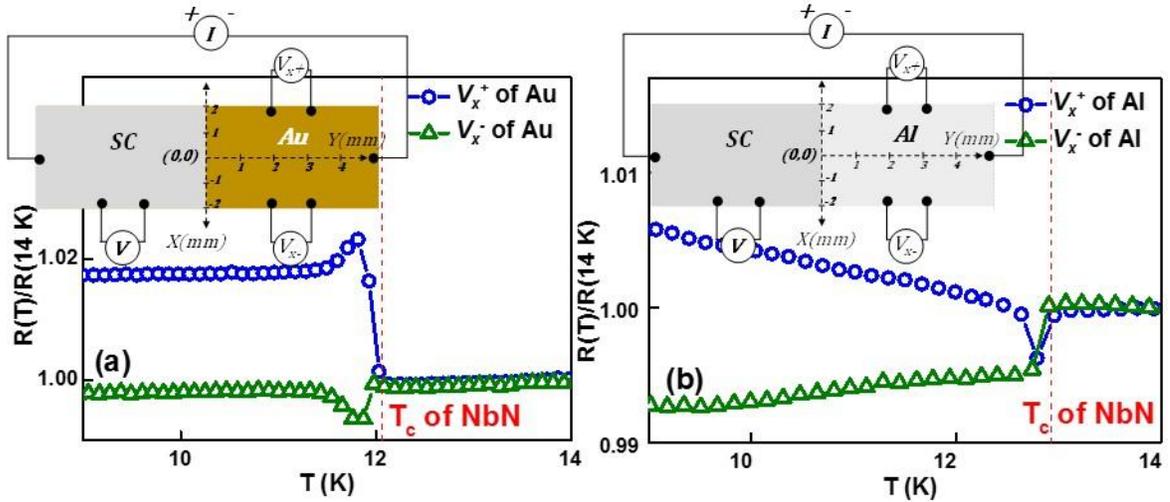

*Figure 6. R-T measurement of (a) bare-Au layer of the NbN-Au heterostructure and (b) bare-Al layer of the NbN-Al heterostructure. Resistance is normalized to the resistance value at 14 K. The red dashed line indicates the superconducting transition temperature ($T_c$) of the NbN superconductor. The heterostructure and the measurement configuration of NbN-Au and NbN-Al heterostructures are added as insets in (a) and (b), respectively.*

The resistance jumps observed in NbN-Al heterostructure is shown in Figure 6 (b). The $T_c$ of the NbN superconductor has been observed to vary up to 1 K during different depositions, possibly due to small changes in the deposition conditions. This causes the difference in the onset of the anomalous resistance jumps in the NbN-Au and NbN-Al heterostructures seen in Figure 6. As was the case in the NbN-BSS heterostructures, increasing magnetic field strength was found to lower the temperature of occurrence of the jump in accordance with the decreasing $T_c$ of NbN in the NbN-Au and NbN-Al heterostructures. Unlike in the case of the NbN-BSS heterostructure, two clear jumps were observed in NbN-Au and NbN-Al heterostructures. The first jump is at the $T_c$ of NbN and the second jump at $T < T_c$. However, as in the case of NbN-BSS, the sign of the net difference between the voltage drop in the superconducting and normal state ($V_S$-$V_N$) is opposite in the $V_x^+$ and $V_x^-$ pairs.

In all the heterostructures represented so far, the $V_x^+$ has showed a net positive jump in resistance and the $V_x^-$, a net negative jump. However, multiple similar heterostructures of NbN-BSS were fabricated and some of these heterostructures showed a net negative jump at $V_x^+$ and a net positive jump at $V_x^-$. These results indicate that the edge along which the upturn occurs is arbitrary and varies from heterostructure to heterostructure. However, if an upturn occurs at one edge of a given heterostructure, the other edge is forced to exhibit a downturn only.

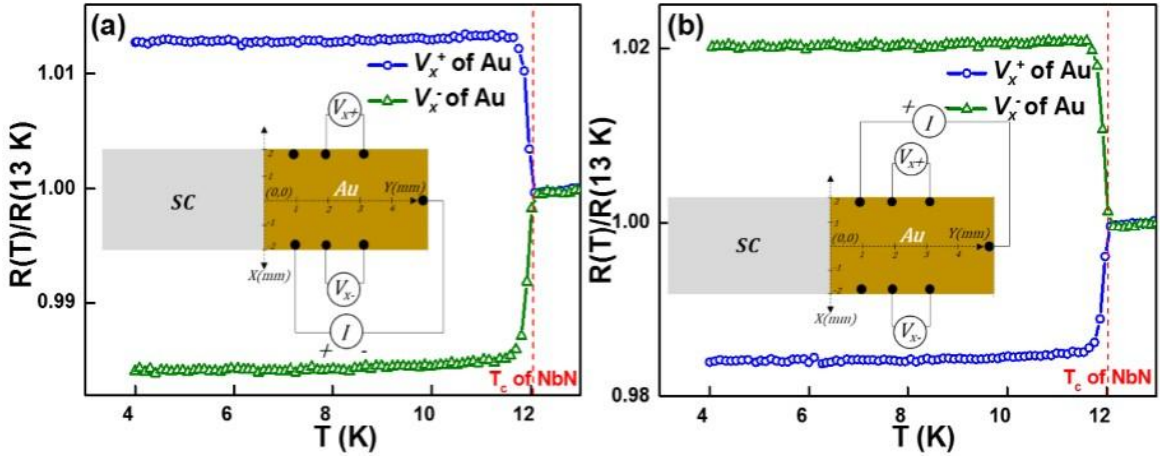

*Figure 7. Two measurement configurations are shown in which the superconductor has been isolated from the current path in the NbN-Au heterostructure. (a) Inset shows the schematic of the measurement configuration in which the current is sourced from a lead on the negative X edge of the bare-Au instead of being sourced from the superconducting layer. The R-T curves at $V_x^+$ and $V_x^-$ pairs for this current configuration are shown. The red dashed line indicates the superconducting transition temperature ($T_c$) of NbN. (b) Inset shows the configuration in which current is sourced from the positive X edge of the bare-Au layer. The R-T curves at $V_x^+$ and $V_x^-$ pairs for this current configuration are shown.*

### 3.1. Influence of current path

In a series circuit, a sudden change in resistance in one part of the circuit can cause non-equilibrium effects in the rest of the circuit. At the superconducting transition temperature the superconductor's resistance abruptly decreases by ~ 6-7 orders of magnitude. Thus to study the possibility that the anomaly originates from such effects, the superconductor was isolated from the current path by sourcing the current from a lead along one of the edges of the bare-TI or bare-metal layer. The two different measurement configurations adopted in the superconductor-isolated circuit of the NbN-Au heterostructure is shown in the insets of Figure 7 (a) and (b). Firstly, it is observed that the anomalous jump in resistance is observed at $T_c$ even when the superconductor is isolated from the current path. Secondly, it is seen that the strength of the upturn/downturn is enhanced in the NbN-Au heterostructure. The opposite sign of the jump on the two edges is retained, however with an important difference.

As depicted in Figure 7 (a), when current is sourced from a lead on the negative $X$ edge of the bare-Au layer $V_x^+$ is seen to exhibit an upturn while $V_x^-$ is seen to exhibit a downturn. But when the current is sourced from a lead on the positive $X$ edge of the bare-Au layer as shown in the inset of Figure 7 (b), $V_x^+$ is found to exhibit a downturn while $V_x^-$ is found to exhibit an upturn. This interesting effect has been observed in the NbN-BSS and NbN-Al heterostructures as well, when the superconductor is isolated from the current path.

### 3.2. Possibility of Hall mixing

Such current path dependent changes in voltage as seen in Figure 7, and the variation in voltage addition along the width of the sample is often characteristic of Hall mixing. Hence, the possibility of the anomalous jump originating from Hall mixing has been thoroughly studied. The Hall voltage changes sign when the current direction or the magnetic field direction is reversed, in accordance with the Lorentz force equation: $F = q(v \times B)$. However, when the current direction was reversed in the configuration where the current is passed across the entire heterostructure (configuration in Figure 2 (b)), no change in the sign or strength of the anomalous jump was observed. Reversal of magnetic field direction also did not produce any change in the anomalous jump, thus suggesting that Hall mixing is not the origin of the anomalous jump.

In any case, the jump has been observed in R-T measurements carried out without the application of any external magnetic field, when in fact Hall effect can take place only in the presence of a perpendicular magnetic field component. To rule out the possibility of a remnant magnetic field in the superconducting magnet acting as the source of Hall effect at zero applied magnetic field, the R-T measurement was carried out in a He Dewar-dipstick setup which does not include a magnet. The anomalous upturn/downturn was observed in the R-T measurements carried out in the Dewar as well. These results conclusively prove that the observed anomaly is not due to the mixing of Hall component.

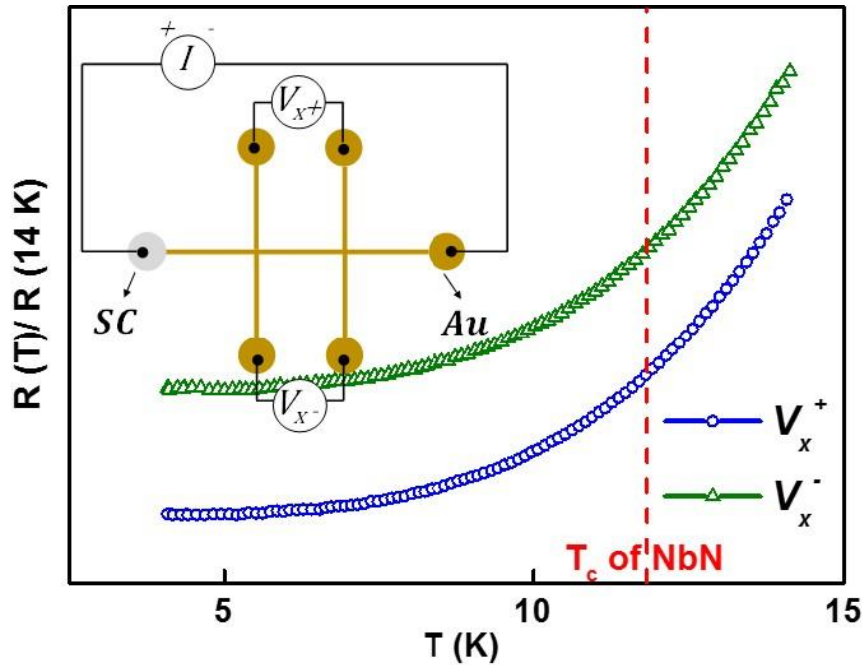

*Figure 8. R-T measurement carried out on the $V_x^+$ and $V_x^-$ voltage pairs of the NbN-Au heterostructure fabricated in the Hall bar geometry. The normalized resistance curve of $V_x^+$ is offset from that of $V_x^-$ for the sake of clear representation. The inset shows the measurement configuration in the Hall bar geometry.*

Surprisingly, R-T measurements carried out in the heterostructure fabricated in the Hall bar geometry as shown in the inset of Figure 8 did not show any anomalous resistance jump. The Au metal was deposited on the substrate with a Hall bar shadow mask. The NbN superconductor was subsequently deposited on one of the current contact pads (on top of the Au layer) by masking the rest of the exposed Hall bar. Since it has already been proved that Hall mixing does not cause the anomaly, it is possible to speculate that the absence of the anomaly in the hall bar geometry is due to the thinner arm of the Hall bar (1 mm) as compared to the width of the heterostructure shown in Figure 1 (4 mm). It is possible that the arm of the Hall bar is not wide enough for the current/voltage inhomogeneity across the width to be perceivable, unlike in the case of the 4 mm wide heterostructures discussed so far. This result hints that the observed anomaly could be due to a current redistribution effect at the $T_c$ of the superconductor.

### 3.3. Current redistribution effect at the superconducting transition

The anomalous jump discussed so far has a clear variation across the width of the sample. This implies that any current/voltage redistribution at $T_c$, if that is the case, must also vary across the width. However, there is no reason for the occurrence of such a width-varying effect in the heterostructures discussed so far (configuration in Figure 2 (b)), wherein the interface is perpendicular to the direction of current flow. A variation along the width can possibly be expected if the interface is not perpendicular to the current direction, but makes an angle θ with the X-axis. Hence, two heterostructures with oblique interfaces were fabricated by placing the shadow mask on the Au layer such that it deliberately made an angle θ and – θ (approximately) with the X-axis. The NbN superconductor was deposited on the region of the Au layer which was exposed by the oblique shadow mask. The schematics of the heterostructures thus fabricated are shown in Figure 9.

The R-T measurements carried out on $V_x^+$ and $V_x^-$ of the two heterostructures are shown in Figure 10. The –θ heterostructure in Figure 10 (a) shows a downturn at $V_x^+$ and an upturn at $V_x^-$. The opposite result is seen in the +θ heterostructure, shown in Figure 10 (b), where an upturn is observed at $V_x^+$ and a downturn at $V_x^-$. In light of these results, the following model is proposed to explain the anomalous resistance jump in the superconductor-metal heterostructures.

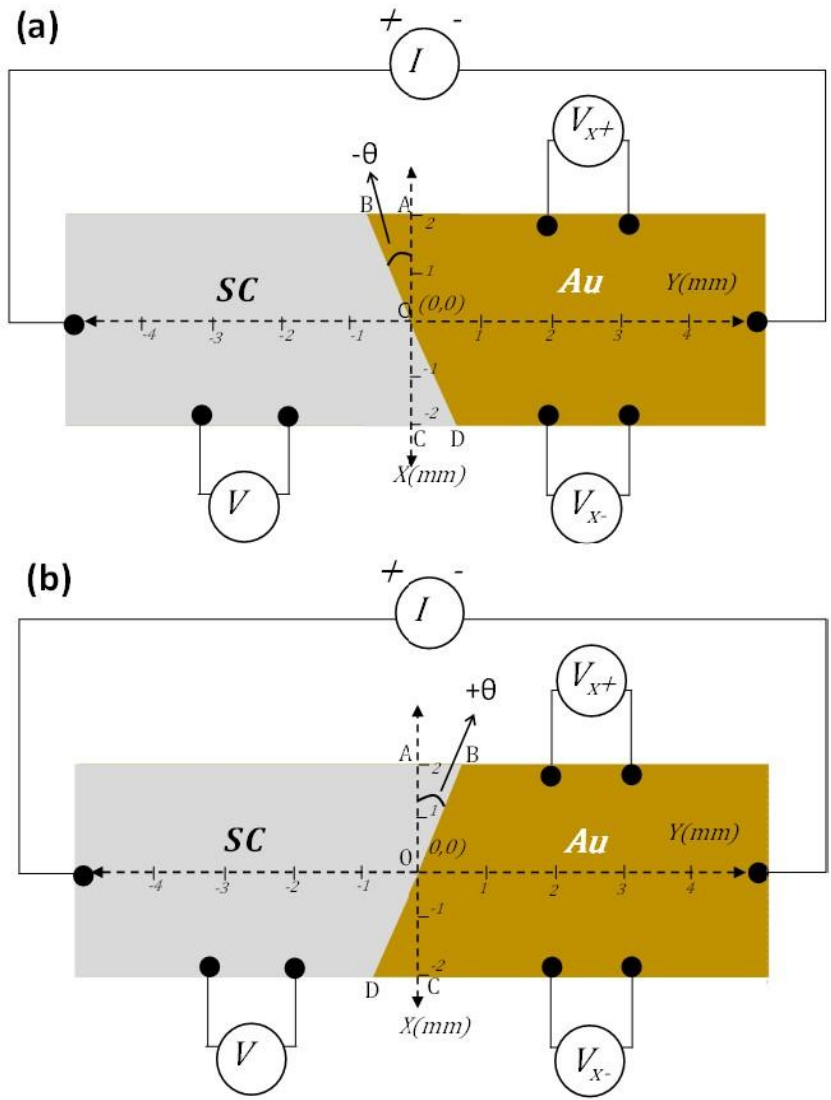

*Figure 9. NbN-Au heterostructures fabricated with oblique interfaces which form angle (a) –θ and (b) +θ with respect to the X-axis. The measurement configuration is also shown.*

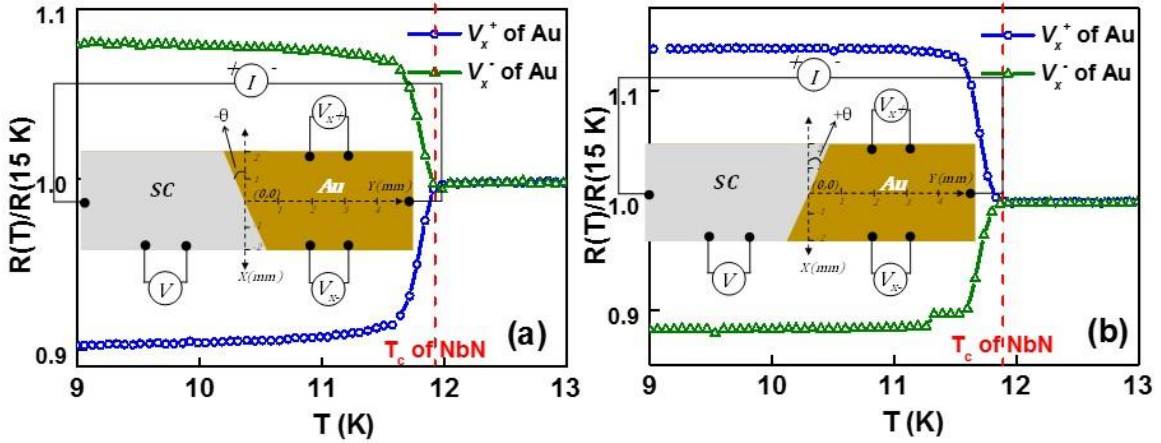

*Figure 10. R-T curves of the $V_x^+$ and $V_x^-$ pair of leads in heterostructures with oblique interface of angle (a) $-\theta$ and (b) $+\theta$. Schematics of the corresponding heterostructures are added as insets of (a) and (b).*

### 3.4. Current redistribution (CRD) model

The model considers the oblique interface heterostructures represented in Figure 9 (a) and (b). In Figure 9 (a), at T > $T_c$, both the NbN superconductor and the Au metal are in the normal state and the order of their conductivities are comparable. Current is uniform across the width of the sample. At T < $T_c$, where NbN is in the superconducting state, the conductivities of NbN and Au differ by ~ $10^{10}$ times. Considering the regions immediately close to the interface in figure 9 (a), $\Delta$ OAB and $\Delta$ OCD are both of comparable conductivities at T > $T_c$. At T < $T_c$, the conductivity of $\Delta$ OCD becomes ~$10^{10}$ times higher than that of $\Delta$ OAB. Due to the preference for the least resistance path, the current entering $\Delta$ OAB curves towards $\Delta$ OCD. This causes an increase in the current density in and around $\Delta$ OCD and there is a corresponding decrease in current density in and around $\Delta$ OAB. The effect of this increased current density in $\Delta$ OCD is felt up to mm length scales ($V_x^-$). The increase in current on the $-X$ arm translates as an increase in voltage drop along this arm due to Ohm's law. Since the resistance value observed through the experiment is obtained by dividing the voltage drop by the constant bias current (1 mA), the sudden increase in voltage at $T_c$ (as the temperature decreases) along the $-X$ arm is reflected as the upturn in resistance in the $V_x^-$ pair of leads. Correspondingly, the decrease in current in the $+X$ arm causes a decrease in voltage due to Ohm's law, which is reflected as a downturn in resistance in the $V_x^+$ pair of leads. There is a gradual variation in the change in the current distribution across the width below $T_c$, and there is expected to be a net zero change in current density along the middle of the width ($X = 0$ line). This is seen in the decreasing amplitude of upturn/downturn as the leads are moved towards the middle of the width ($X = 0$ line) in Figure 5 (b) and an almost vanishing amplitude of jump in the *M* pair of leads.

In the heterostructure with +θ interface angle with respect to the *X*-axis (Figure 9 (b)), Δ OAB goes into the superconducting state and its conductivity increases ~ $10^{10}$ times compared to that of Δ OCD, below $T_c$. Hence at $T < T_c$ the current entering Δ OCD curves towards Δ OAB, as it is the smaller resistance path by several orders of magnitude. Current density in and around Δ OAB increases and current density in and around Δ OCD decreases. This causes an increase in the voltage drop along the +*X* arm and a decrease in the voltage drop along the –*X* arm, causing an upturn in $V_x^+$ and a downturn in $V_x^-$. Hence, if the sign of the angle of the interface with respect to the *X*-axis changes, the sign of the upturn/downturn anomaly also changes.

It is to be noted that the interface is a step and hence the 2D model discussed here is a simplified picture. However, the experiments were repeated in a lateral geometry where the superconductor and metal were deposited side by side and qualitatively similar results were obtained.

### 3.4.1. Dependence on the angle of the interface

In the heterostructures studied in Figure 1 to Figure 8, the interface is intended to be exactly perpendicular to the current direction which is along the *Y*-axis, when sufficiently away from the current leads. That is, the interface is intended to be exactly aligned with the *X*-axis. However, since the positioning and securing of the shadow mask is carried out manually on hand-cut Si/SiO$_2$ substrates, it is entirely plausible that the interface is not exactly aligned with the *X*-axis. In fact, the probability of the interface being perfectly aligned with the *X*-axis tends to zero. On closer inspection of the films studied and represented in Figure 1 to Figure 8, it was found that the obliqueness of the interface angle was not perceivable to the naked eye up to ±5°. For instance, the interface angle was found to be ~ +5° (found under the microscope) for the NbN-Au heterostructure, the R-T results of which are presented in Figure 6 (a). But this obliqueness of the interface was barely noticeable to the naked eye. The angle of the heterostructure with the deliberately oblique interface shown in Figure 10 (b) is ~ +50°. The percentage of change in voltage drop (ΔV%) of the $V_x^+$ upturn in the +5° heterostructure of Figure 6 (a) is found to be ~1.2%, while that of the +50° heterostructure of Figure 10 (b) is found to be ~14%. That is, when the angle of the interface is ten times greater, the percentage of change in the voltage jump is also ~10 times greater. There is a direct relationship between ΔV% and the angle of the interface with respect to the *X*-axis, for given materials and dimensions. This implies that an angle as small as 0.5° can cause a 0.1% change in ΔV.

It should also be noted that the apparent arbitrariness in the edge along which the upturn occurs (downturn occurs on the other edge) is in fact not arbitrary at all. Rather it depends on the sign of the angle of the interface with respect to the *X*-axis. A –θ angle of interface due to an obliqueness in the positioning of the shadow mask causes a downturn along $V_x^+$ (upturn along $V_x^-$) and a +θ angle of interface causes an upturn along $V_x^+$ (downturn along $V_x^-$). The reasons are as discussed in the CRD model.

### 3.4.2. Dependence on the bias current magnitude

The current redistribution model (CRD model) proposes that the anomalous resistance jump is due to an asymmetric distribution of current across the width of the film below $T_c$, where one edge receives more current while the other edge experiences less current. Hence according to this model, the increase in current and hence the voltage drop along one edge (or the decrease along the other) should be proportional to the constant current that the heterostructure is biased with. In other words, the percentage of change in the voltage drop or the ΔV% must remain the same for different constant bias current values for currents that do not destroy the superconductivity of NbN. This has been observed experimentally in an NbN-BSS heterostructure by measuring the R-T curves for different bias current values, as shown in Figure 11. The Voltage drop-temperature curves of $V_x^+$ and $V_x^-$ are added in the inset of Figure 11 (a) and (b), respectively. This observation further confirms the validity of the CRD model that has been proposed to explain the anomalous jump in resistance.

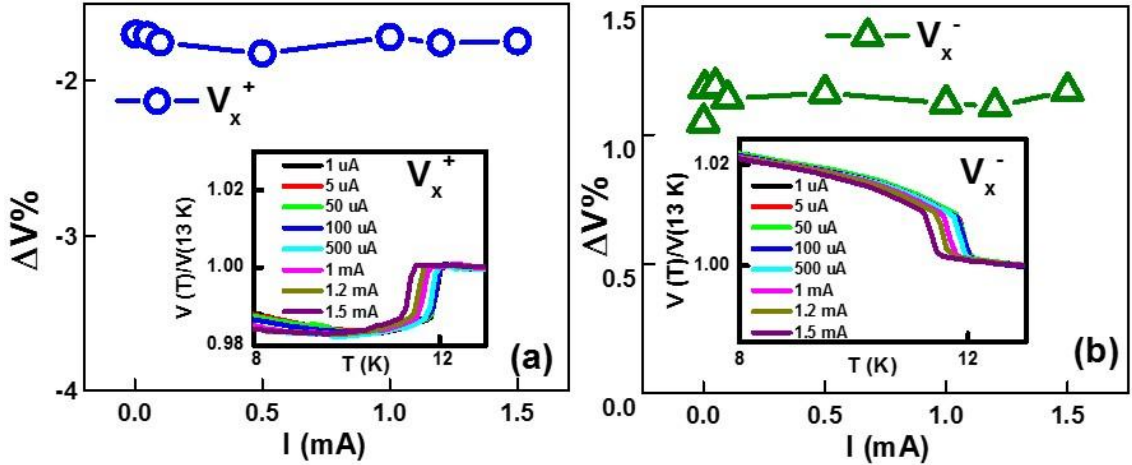

*Figure 11. The percentage of change in voltage drop (ΔV%) at the anomalous jump in NbN-BSS heterostructure for different constant bias current values along (a) $V_x^+$ and (b) $V_x^-$. The insets in (a) and (b) show the V-T curves at different constant bias current values along $V_x^+$ and $V_x^-$, respectively, from which the ΔV% values have been estimated.*

### 3.4.3. Variation with distance from interface

The current redistribution due to the large asymmetry in conductivity along the width, is expected to be strongest near the interface. As the separation from the interface increases it is expected that the current becomes increasingly homogenous across the width. This implies that the strength of the anomalous jump should decrease as the distance from the interface increases. A long NbN-Au heterostructure of length 12 mm and width 4 mm was fabricated to study the anomalous jump as a function of distance from the interface. The NbN superconductor was deposited on the Au layer over a length (*Y*-axis) of 2 mm and a width of 4 mm, as shown in the schematic in Figure 12 (a). Equally spaced voltage leads were fabricated along the +*X* and –*X* edges of the bare Au region of the heterostructure. R-T measurements were carried out in the electrical configuration shown in the schematic and the results are depicted in Figure 12 (b) and (c). As can be seen from the figures, it was observed that the amplitude of the upturn and downturn decreases with increasing separation from the interface, which is in agreement with the proposed model. It is to be noted that the interface in the given figure appears to be aligned with the *X*-axis, but in the actual experimental heterostructure it presumably possessed a small angle with respect to the *X*-axis, in accordance with the CRD model.

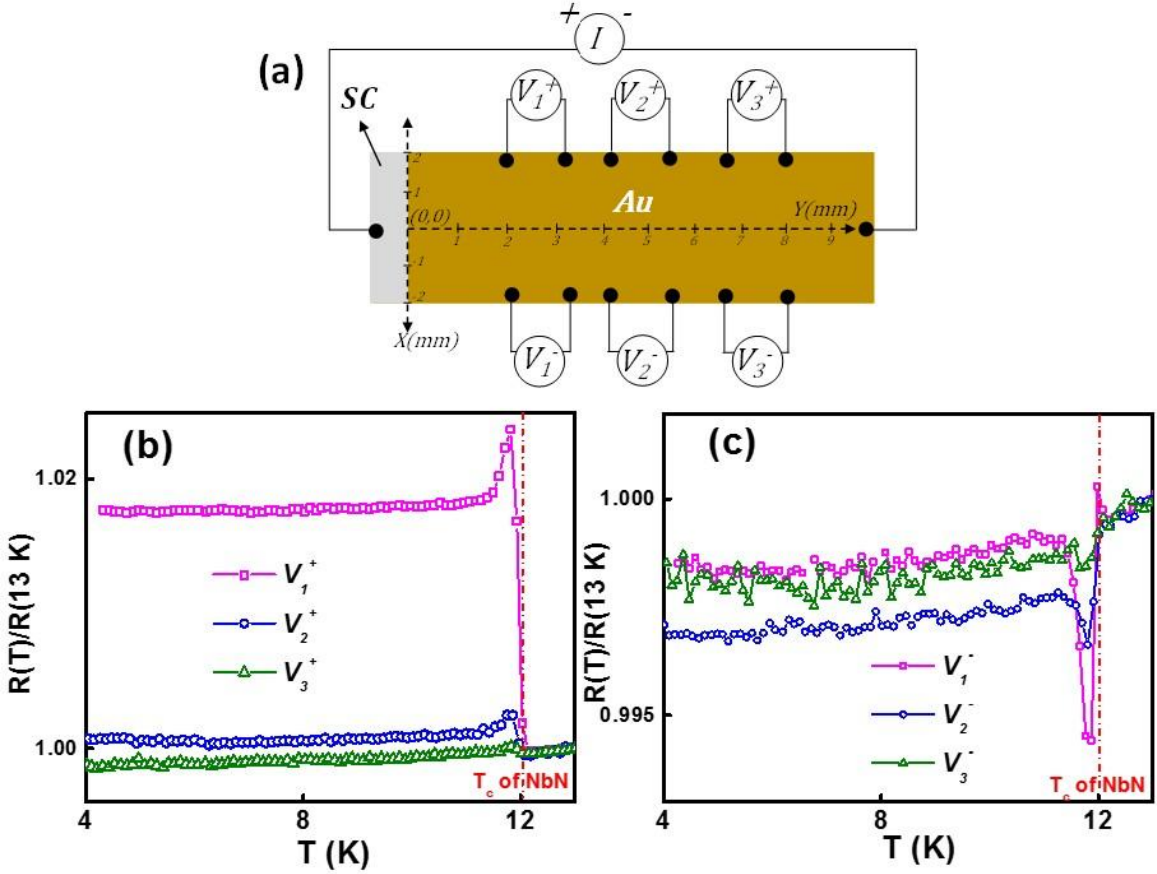

*Figure 12. (a) Schematic of a long NbN-Au heterostructure with leads along $X^+$ and $X^-$ edges at different distances from the interface. R-T measurements on the equally spaced voltage leads along the (b) +X edge and the (c) –X edge depicting the variation in the anomalous resistance upturn/downturn as a function of separation from the interface.*

### 3.5. Finite element analysis using COMSOL

The model proposed to explain the anomalous resistance behaviour invokes the simple physical phenomenon of Ohm's law. This should hence be verifiable by using a simulation technique, which allows the user to choose the physical equations that can be involved in the simulation. One such powerful tool is finite element method based multi-physics solver COMSOL. COMSOL version 5.3a has been used to carry out the simulations presented in this study.

The CRD proposed in section 3.4 is based solely on the vast difference in conductivity between the metal and the superconductor. Hence the superconducting layer has been simulated in the finite element method (FEM)-based model by invoking only the superconductor's property of highly enhanced conductivity below $T_c$. The model would be more accurate if the conduction in the superconducting layer were only up to the penetration depth, as is the case when the applied magnetic field is less than the lower critical magnetic field. However, the penetration depth value is in the range of a few nm at $T_c$, which poses the problem of a high aspect ratio while meshing. Hence in the FEM model, the entire superconducting layer is considered to possess very high conductivity ($\sim 10^{14}$ S/m), modeling the current to flow throughout the material. In any case, this rudimentary model of the superconductor has been able to qualitatively mimic the obtained experimental results, as shown in the discussions that follow. The physics of *electric currents*, which is available under the *AC/DC* module of the COMSOL package has been invoked in the model. The physical concepts involved in the *electric currents* module are the Ohm's law and the law of conservation of current. The equations that are part of the *electric currents* module are as follows.

*Continuity equation:* $\quad \nabla \cdot J = -\partial \rho / \partial t \quad$ -Equation 1

*Ohm's law:* $\quad J = \sigma E \quad$ -Equation 2

$$E = -\nabla V \quad \text{-Equation 3}$$

$J$ is the volume current density, $\rho$ is the volume charge density, $t$ is time, $E$ and $V$ are the electric field and electric potential at a given point, respectively. A constant bias current is set between current leads and the model primarily solves for the electric potential $V$ at all points within the computational domain.

### 3.5.1. Computational Model

The basic model consists of two cuboidal layers, similar to the thin film heterostructure. Layer 1 is built such that it is 4mm wide and 10 mm long, while layer two rests on layer 1 and has a width of 4 mm and a length of 5 mm. Figure 13 (a) gives a clear picture of the basic model. Layer 1 is equivalent to the TI or metal layer of the experimental heterostructure and layer 2 is equivalent to the superconductor. While the thickness of each of the two layers in the experimental heterostructure is ~100 nm, due to the problem faced during meshing of models with large aspect ratio, the thickness of the two layers in the FEM model has been chosen to be one order higher (1 μm). The model shown in Figure 13 (b) mimics the ideal scenario where the interface angle θ is equal to 0°. However several models were built to depict the real scenario of shadow masking where the interface angle can have a non-zero interface angle of –θ and +θ, respectively. This is achieved by replacing the cuboidal layer 2 with a hexahedron of appropriate dimensions.

A constant bias current of +1 mA is set across the model, between two cylindrical current leads, which mimic the silver paint contacted Copper leads in the experimental heterostructure. The electrical conductivity (σ) of the copper leads is set to be $10^7$ S/m, σ of layer 1 (metal/TI layer) is fixed to be $10^5$ S/m. The normal state (T > $T_c$) scenario has been simulated by setting the σ of layer 2 (superconducting layer) to $10^5$ S/m and the superconducting state (T < $T_c$) has been simulated by setting the σ of layer 2 to $10^{14}$ S/m, which is $10^9$ times higher than the conductivity of layer 1. The normal state conductivity of $10^5$ S/m has been chosen to mimic the conductivity values of NbN and $Bi_2Se_3$ thin films. The volume current density distribution has been simulated in the normal and pseudo-superconducting state scenarios and have been compared with each other. The electric voltage values at different points have also been obtained from the computation and the voltage difference behaviour at different regions of the bare metal layer has been compared with the experimental results obtained in the normal and superconducting states. These comparisons have been performed for different interface angles to bring out the role played by the interface angle in redistribution of current. The number of degrees of freedom in the finite element meshing process was seen to be ~ 35000 for all simulated geometries.

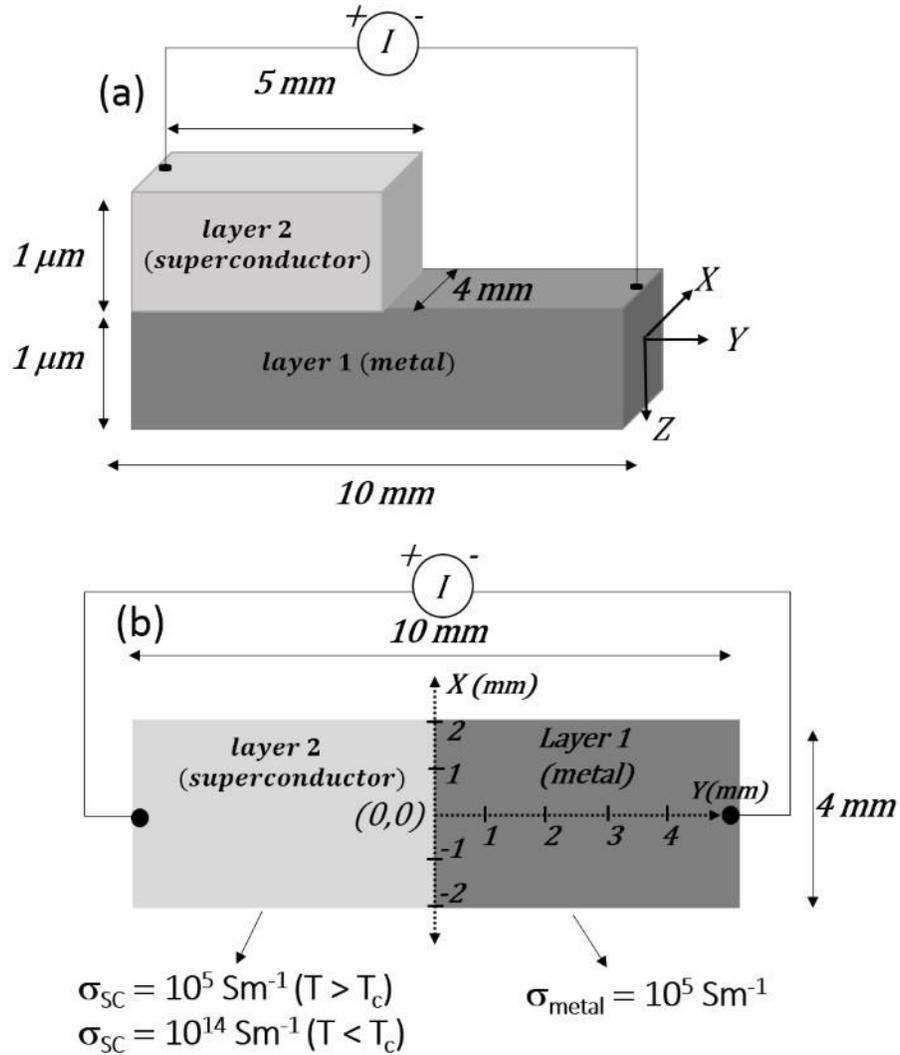

*Figure 13. (a) The lateral view and (b) top view of the superconductor-metal heterostructure model used in COMSOL simulations.*

### 3.5.2. Results of finite element analysis
#### 3.5.2.1. Interface angle dependence

In models with a non-zero interface angle θ, an asymmetry is observed in the volume current density across the width in the normal state (T > $T_c$). However, this asymmetry becomes much stronger in the superconducting state. Figure 14 (a), (b), (c), (d), and (e) depict the volume current density simulation of the model in the normal state (T > $T_c$) and the superconducting state (T < $T_c$) for interface angles of 0°, -5°, +5°, -20°, +20°, respectively. In the case of 0° angle, shown in Figure 14 (a), there is no perceivable change in the current density along the *X*-axis. However, in the case of -5° and -20° simulations shown in Figure 14 (b) and (d), at T < $T_c$ a higher current density is seen in the (-*X, +Y*) region as compared to the (+*X, +Y*) region. On the other hand, in the case of the +5° and +20° models, shown in Figure 14 (c) and (e), a higher current density is seen in the (+*X, +Y*) region as compared to the (-*X, +Y*) region. This asymmetry along the *X*-axis is seen to be strongest near the interface and fades away with increasing distance from the interface. A comparison of the -5° and -20° current distribution (Figure 14 (b) and (d)) shows that the strength of the asymmetry becomes stronger with increasing angle of the interface. These results confirm the CRD model proposed in section 3.4.

The voltage difference across pairs of points 1 mm apart at increasing separations from the interface and at different regions across the width of the metal layer have been obtained from the computations for models with interface angle θ = 0°, ±5°, ±10°, and ±20°. The voltage difference values were obtained in the normal state and in the superconducting state and the percentage of anomalous voltage jump was obtained using $\Delta V\% = \frac{V_S - V_N}{V_N} \times 100$
- Equation 4, where $\Delta V\%$ is the percentage of anomalous voltage jump, $V_S$ is the voltage difference across a given pair of points 1 mm apart in the superconducting state, and $V_N$ is the voltage difference across a given pair of points 1 mm apart in the normal state.

$$\Delta V\% = \frac{V_S - V_N}{V_N} \times 100 \qquad \qquad \textit{- Equation 4}$$

The $\Delta V\%$ has been plotted as a function of the position along the width (position along *X*-axis) for models with different interface angles. Figure 15 (a), (b), and (c) represent such plots at separations of 1 mm, 2 mm, and 3 mm from the *X*-axis (distance between the first point in the pair and the *X*-axis). In models with interface angle of -θ, the $\Delta V\%$ is negative for positive *X* values indicating a decrease in $\Delta V$ along the +*X* edge as system goes into the superconducting state. This is the downturn scenario observed in the R-T experiments. In models with interface angle of +θ, the $\Delta V\%$ is positive for positive *X* values, indicating an increase in $\Delta V$ along the +*X* edge as system goes into the superconducting state. This is the upturn scenario observed in the R-T experiments. A non-linear $\Delta V\%$ variation is observed across the width. The slope of the curve increases with increasing θ. The value of $\Delta V\%$ observed in the ±5° models are close to the values observed in the NbN-BSS experimental heterostructures.

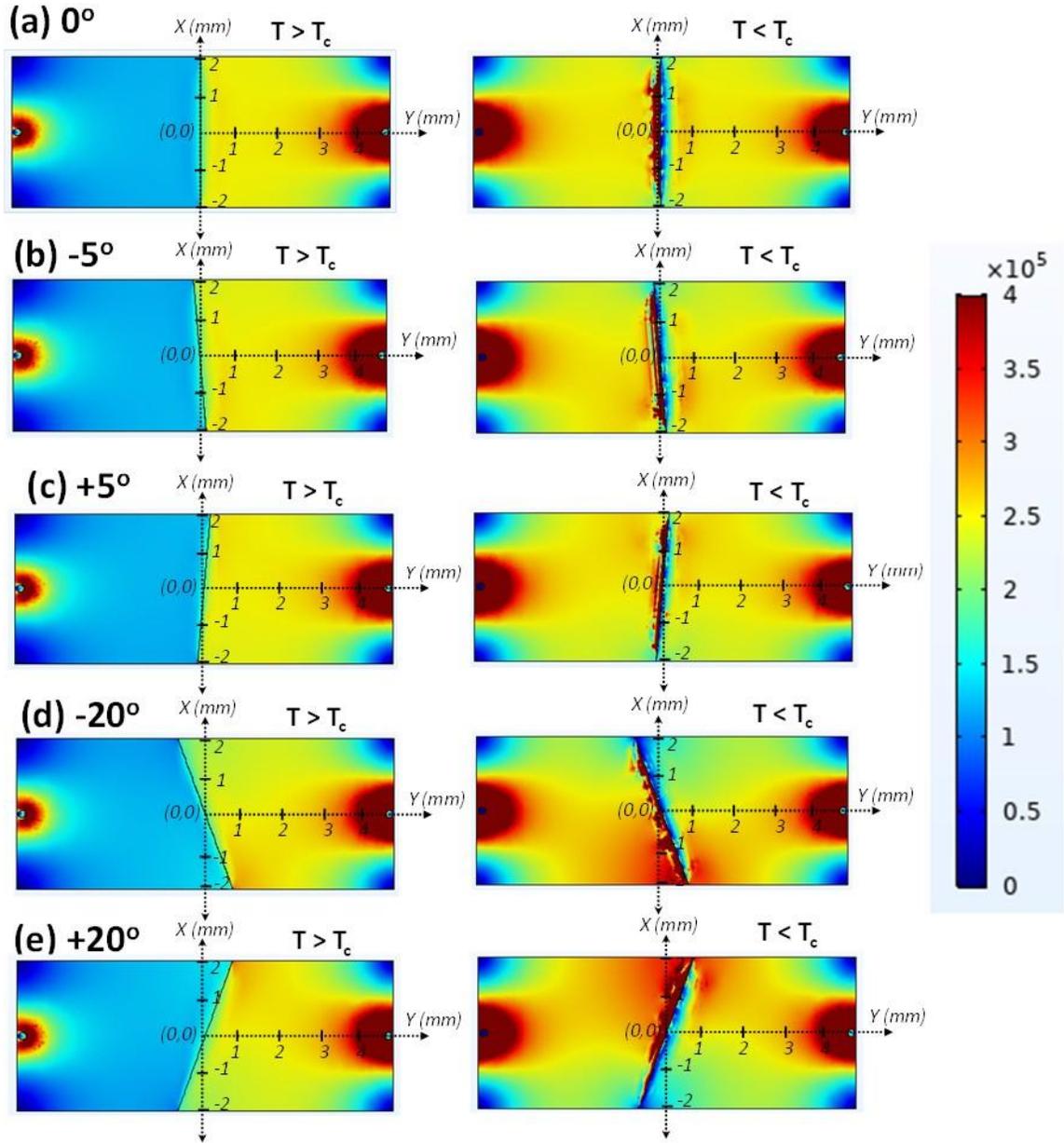

*Figure 14. The volume current density distribution simulated using COMSOL in the normal state ($T > T_c$) and the superconducting state ($T < T_c$) in a superconductor-metal heterostructure model with interface angle (a) $\theta = 0°$, (b) $\theta = -5°$, (c) $\theta = +5°$, (d) $\theta = -20°$, and (e) $\theta = +20°$.*

A comparison of $\Delta V\%$ in Figure 15 (a), (b), and (c) shows that the anomalous jump becomes weaker with increasing separation from the interface. A colour map representation of the variation in $\Delta V\%$ across the surface of the metal layer of the model with $\theta = -5°$ is shown in Figure 15 (d).

Figure 15 (d) clearly captures the experimental results of strong upturn/downturn along the negative/positive X-coordinates near the interface, which becomes weaker with increasing separation from the interface. $\Delta V\%$ at a given position is found to vary linearly as a function of the interface angle θ, as shown in Figure 15 (e).

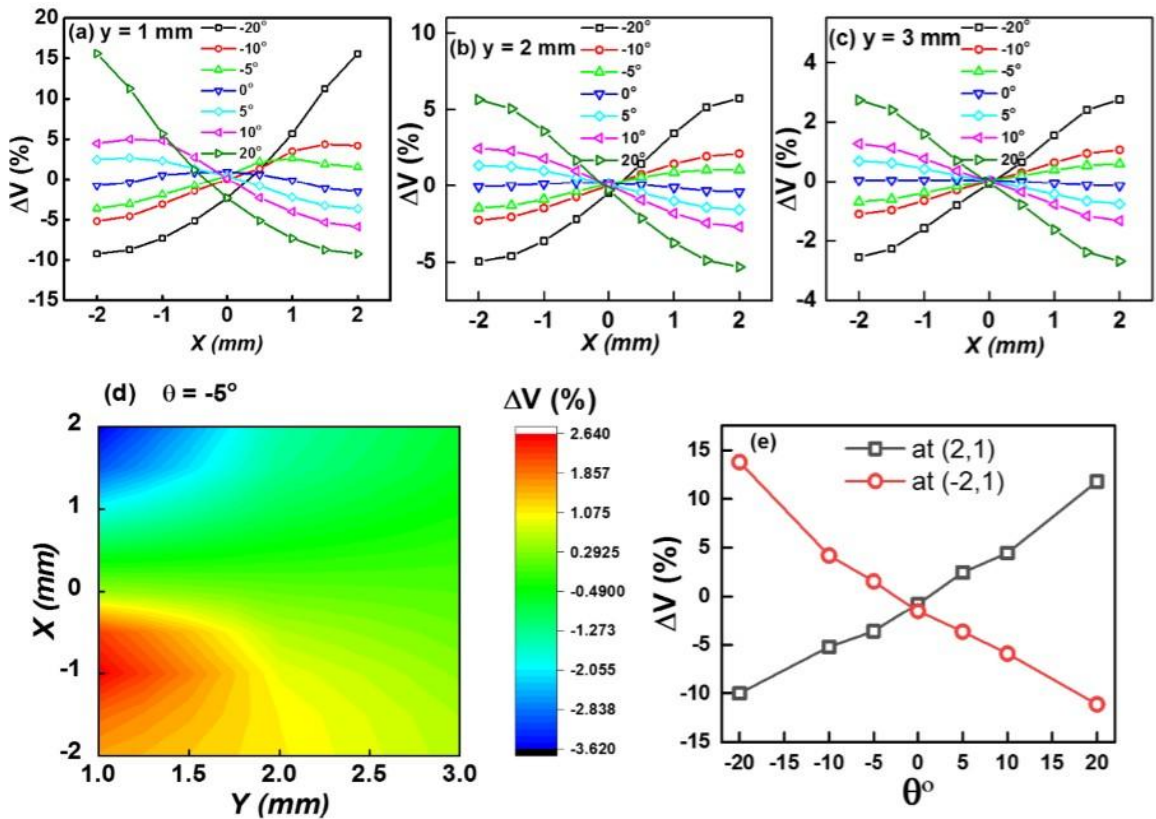

*Figure 15. The anomalous voltage jump percentage ($\Delta V$ %) plotted across the width of the bare metal layer (along the X-axis) at (a) 1 mm, (b) 2 mm, and (c) 3 mm separations from the X-axis for metal-superconductor heterostructures with different interface angles θ. It is to be noted that all voltage drop values are calculated between a pair of points 1 mm apart from each other along the Y-axis. (d) Colour map representation of the variation of $\Delta V$ % along the X-axis and the Y-axis of the bare metal layer. (e) The linear variation of $\Delta V$ % at specific regions on the bare metal layer as a function of the interface angle θ. The $\Delta V\%$ is obtained through FEM simulation.*

### 3.5.2.2. Current path dependence

It was shown experimentally (see Figure 7) that shifting the current source to either edge of the metal layer, attempting to isolate the superconductor from the current path, drastically changes the upturn/downturn results obtained. In order to understand these results and the possible influence of interface angle in this scenario, the basic FEM model discussed in section 3.5.1 was modified such that the current source lead is on either edge of the metal (along +X or -X), as shown in Figure 16. A study of the volume current density distribution in the normal and superconducting states of model with $\theta = 0°$ showed the following. When the current source lead is along the +X edge and when system goes into the superconducting state, current is shared by the superconductor near the +X edge (redistributed into the superconductor near the +X edge), causing a decrease in current density in +X edge, and hence causing a decrease in V when the system goes to the superconducting state, i.e., downturn in R-T when system cools to the superconducting state. The opposite is observed when current is sourced from the -X edge. Hence, the anomalous voltage jump is observed even in a perfectly $\theta = 0°$ heterostructure if the current is not biased symmetrically across the interface. This is due to current sharing by the region of the superconductor near the current source. As seen from Figure 16 (c), the interface angle does not affect the jump strongly. The -20° model with current sourced from -X edge alone shows an enhanced $\Delta V\%$ variation, possibly because the current lead in this case is very close to the superconducting interface, causing a stronger sharing/redistribution of current.

### 3.5.2.3. Conductivity dependence

The experimental studies have involved measurements on NbN-Au and NbN-Al heterostructures in addition to the NbN-BSS heterostructures modelled using FEM so far. In order to simulate the NbN-Au heterostructure, the conductivity of the metal layer (layer 1) was changed to $10^8$ S/m. The $\Delta V\%$ of the $\theta = -5°$ NbN-Au model at any given region was found to be higher than that of the NbN-BSS model. In order to obtain a general picture of the effect of material conductivities on the anomalous voltage behaviour, the computation was performed on $\theta = -5°$ models of varying conductivity values of the metal layer and the superconductor layer (normal state conductivity).

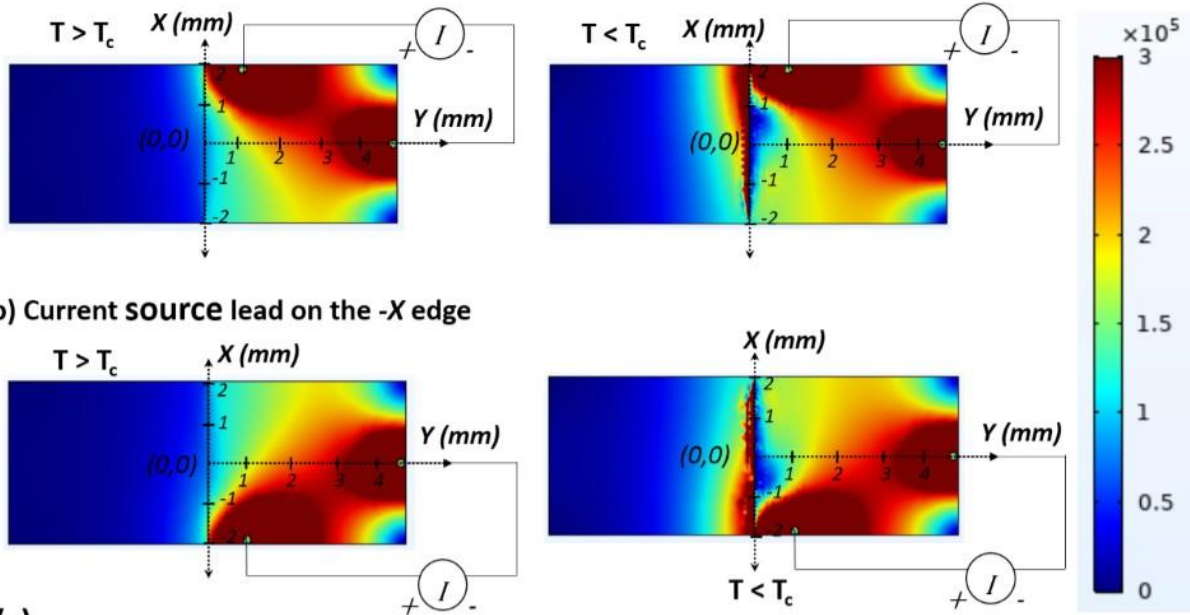
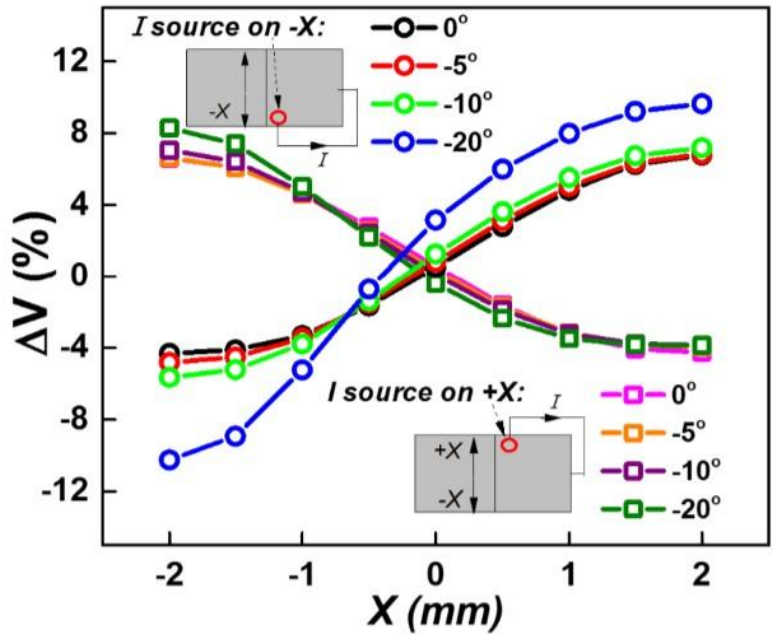

*Figure 16 (a) The volume current density distribution simulated using COMSOL in the normal state ($T > T_c$) and the superconducting state ($T < T_c$) in a superconductor-metal heterostructure model, where the source current lead is positioned on the (a) +X-edge and on the (b) -X-edge. (c) A comparison of the ΔV% across the width of the bare metal layer of heterostructures with different interface angles for two different current source configurations: I source on the +X-edge or the -X-edge. The ΔV% is obtained through COMSOL simulation.*

A colour map representation of the results is shown in **Error! Reference source not found.**Figure 17. It is evident that the current redistribution (and hence the anomalous jump) is stronger when the conductivity of the metal layer is higher. However, the maximum ΔV% achieved for the highest simulated conductivity ($10^8$ S/m) is only ~ 5%. The anomalous jump is negligible and hence unobservable in heterostructures where the metal layer is insulating (σ < 50 S/m). It is also observed that when the normal state conductivity value of the superconducting layer is comparable to that of the metal, the strength of the anomalous jump is higher.

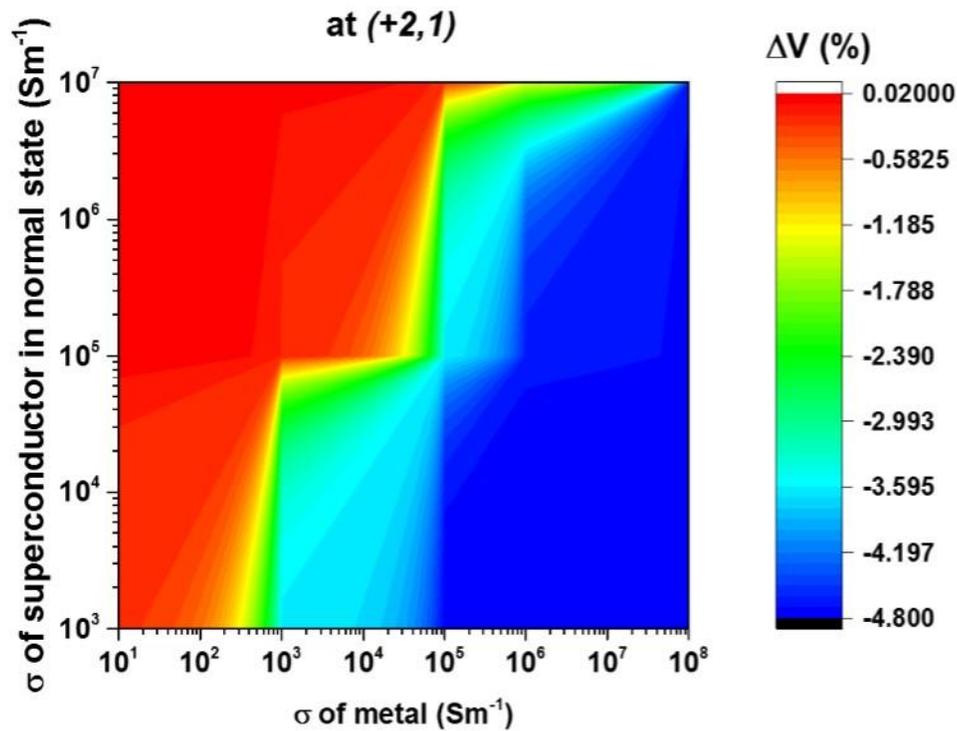

*Figure 17. Colour map representation of the anomalous voltage jump percentage (ΔV%) as a function of the conductivity of superconducting layer in $T > T_c$ (normal state) and the conductivity of the metal layer at a given region (+2,1) on the bare metal layer of a superconductor-metal heterostructure. The ΔV% is obtained through COMSOL simulation.*

The simulated results obtained through FEM are in complete agreement with the CRD model that has been proposed to be the origin of the anomalous upturn and downturn in temperature-dependent and magnetic field dependent resistance measurements in the superconductor-topological insulator and superconductor-metal heterostructures discussed in this study.

### 3.6. Discussion

A sudden and large change in the resistance of one region of a heterostructure can cause the current to redistribute in the entire structure due to preference of current for the least resistive path, in accordance with Ohm's law. As seen from the results in this study, the current redistribution effect can cause interesting observations that can be misinterpreted to have more involved origins, for instance, quantum confinement, spin scattering, or proximity. It is to be noted that the effects due to current redistribution are strong in terms of the length scales up to which they can be observed (~ mm in this study). Since superconducting transition is one of the few transitions in physics that is accompanied by a drastic alteration of electrical and magnetic properties of a material (~$10^{10}$ times increase in electrical conductivity, several order decrease in magnetic permeability), these effects must always be anticipated in superconductor heterostructures. However, the appearance of this artefact may not be limited to superconducting heterostructures. The sharp decrease in electrical resistance at the paramagnetic to ferromagnetic transition may also cause modifications in current path which could appear as resistance upturn/downturn near a ferromagnetic-metal interface, which could potentially be misinterpreted as long range magnetic effects due to proximity. Inexplicable upturn and downturn (< 2%) in resistances in different pairs of leads have also been observed in high-pressure low temperature resistivity studies on $Bi_{1-x}Sb_xSe_3$. In light of the knowledge of current redistribution at $T_c$, it is now speculated that superconductivity was being induced randomly at different regions of the sample, which is expected [35,36]. This change brought about in the resistance distribution on the surface of the material caused an alteration in the current path, leading to upturn in some pairs of leads (increased current density) and downturn in other pairs of leads (decreased current density). These results are yet to be published.

A number of studies on superconductor-TI heterostructures, attempting to study proximity-induced superconductivity in TI, have adopted the approach where the TI layer is contacted with superconducting electrodes. This interface is similar to the interface in the step geometry discussed in this study. Some of the studies[6,8,19,20,25,29] have reported resistance downturn in the TI at the $T_c$ of the superconductor, attributing it to the induced proximity effect that is observed in any superconductor-metal heterostructures with a sufficiently transparent interface. On the other hand, few studies [20,26–28] have reported an intriguing upturn in the resistance of the TI near the superconductor-TI interface at $T_c$. Wang *et al.*[26] attributed the upturn observed in In-Bi$_2$Se$_3$-In, Al-Bi$_2$Se$_3$-Al, and W-Bi$_2$Se$_3$-W heterostructures to spin-flip transitions at the interface. The study argues that the spin-polarized electronic state of TI surface is incompatible with the spin-singlet cooper-pairing of electrons in *s*-wave superconductors. Hence cooper pairs leaking into the TI at the superconductor-TI interface or spin-polarized electrons leaking into the superconductor at the TI-superconductor interface must undergo spin-flip, which leads to the increase in resistance at $T_c$. This idea gained attention and other studies subsequently attributed resistance upturns to spin-flip transitions. However, it is seen that there could be room for ambiguity in some reports and that it is possible to interpret their results from the point of view of the current re-distribution effects discussed in this study.

Afzal *et al.*[27] have reported a ~1% increase in resistance at the $T_c$ of Indium (In) in an In-MoTe$_2$-In heterostructure, where the In superconducting leads are hand-contacted on MoTe$_2$ TI pellets and the R-T is recorded in the four-probe geometry. The ~1% upturn has been attributed to spin-flip transitions. Firstly, since the In leads are hand-contacted, it is expected that the leads are asymmetrically positioned and irregularly shaped and hence may lead to an enhanced current asymmetry across the sample when the In leads go into the superconducting transition, causing the observed upturn. Secondly, the resistance upturn due to spin-flip reported by Wang *et al.*[26] is more than 100%, while the jump is only 1% in the case of Afzal *et al.*[27], which is more comparable to the ΔV% values observed due to current redistribution effects in this study. However, it cannot be ruled out that the small jump is due to poor transparency of the interface. In a previous study reported by our group on Mo-BSS heterostructures [28] a ~ 1% upturn in the resistance of BSS single crystal flakes at ~ mm separations from the interface was observed and interpreted as being due to the antagonistic ground states of superconducting Mo and spin-polarized TI surface, in accordance with Wang's picture [26]. The Mo-BSS device has been grown in a step geometry using Al shadow mask similar to the growth technique carried out in this study. In light of the findings reported in the present manuscript, it is now understood that the upturn observed in Mo-BSS is due to current redistribution effects, in accordance with the CRD model.

As shown in the present study, the effects of asymmetry in superconducting interface in relation to the current path can also lead to the observation of a downturn at $T_c$ in resistivity studies. *Wang et al.*[20] have reported a double transition in electrical resistivity measurements as a function of temperature and magnetic field in Bi$_2$Te$_3$ exfoliated flakes with hand-contacted In current and voltage leads. While the second transition has been attributed to induced proximity in Bi$_2$Te$_3$, the first transition has been attributed to shunting at the In voltage leads at the T$_c$ of In. However, it is interesting to note that among the three studied samples, the R-T curves of two of them show a resistance downturn, while the R-T curve of one of them shows an upturn. All the jumps are within 1%-4% in amplitude. The SEM images show that the hand-contacted In leads on the Bi$_2$Te$_3$ flakes are irregularly shaped (hence the interface is irregularly shaped). Hence it is plausible that the observed upturn or downturn at T$_c$ is due to current redistribution effects. Recently, Yadav *et al.* [25] reported an 8.4% drop in the resistance across W leads (1.38 μm separation) contacted on Bi$_2$Se$_3$ flake at the T$_c$ of W and Zhang *et al.* [19] observed a much larger ~ 20% decrease in resistance across the junction of a Bi$_2$Se$_3$/Nb heterostructure. In both the reports the resistance drop is speculated to arise from an induced long-range superconducting proximity effect. However, the studies have not been able to ascertain the origin.

It would appear that the solution to avoid the effects of current redistribution at the superconducting transition is careful mechanized patterning of heterostructures. However, as has been shown in section 3.5.2.2, current redistribution at T$_c$ can also be caused due to improper positioning of the current leads such that the current path is asymmetric across the interface. As a result, it is important that the studies reporting sudden and unexpected changes in resistance in superconductor-topological insulator heterostructures and attributing quantum mechanical origins to them should be revisited from the point of view of the classical current redistribution effects.

### 3.7. Conclusion

Superconductor-topological insulator (NbN-Sb-doped Bi$_2$Se$_3$) heterostructures were fabricated in the step geometry via PLD and magnetron sputter techniques with the help of shadow masking to pattern the heterostructure, with the motivation of studying the effect of the superconducting transition in the bilayer on the bare topological insulator region near the interface. Temperature and magnetic field dependent resistance measurements showed an anomalous jump in the electrical resistance behaviour of the bare TI layer at the superconducting transition of the bilayer. This anomalous jump was found to vary gradually across the width of the TI, such that the two opposite edges showed anomalous jumps of opposite sign. It was verified that the phenomenon is not unique to TI and can also be observed in ordinary superconductor-metal heterostructures. The possibility of Hall voltage mixing was ruled out through careful experimentation. It was found that the width-varying anomalous resistance jump originated from the non-zero interface angle made by manual shadow masking. A model has been proposed which attributes the anomaly to current redistribution effects due to the oblique interface. Finite element analysis based simulations have been performed using COMSOL software to confirm the validity of the model. The superconducting state was modelled by attributing a conductivity value of ~$10^{14}$ S/m. The simulations showed that with increasing obliqueness of the interface angle, the strength of current redistribution becomes stronger and also yielded stronger jumps in voltage differences on the bare metal region. It was also seen that the anomaly can appear in superconductor-metal heterostructures even for zero angle of interface, if the current lead is positioned (or shaped) such that the current path is asymmetric with respect to the interface. This study shows that in a heterostructure where the electrical conductivity of one part of the structure changes abruptly and drastically, anomalous artefacts can be detected at ~ mm length scales away from the interface due to redistribution of current which can be misinterpreted to have origins elsewhere.

## Acknowledgement

One of the authors (Abhirami S) would like to thank the Department of Atomic Energy, government of India for financial support.